\documentclass[aps,showpacs,nofootinbib,twocolumn]{revtex4}

\usepackage{graphicx}
\usepackage{bm}
\usepackage{amsmath}
\usepackage{amssymb}
\usepackage{hyperref}
\usepackage{epsfig}

\newcommand{\nslash}{\kern 0.2 em n\kern -0.50em /}
\newcommand{\kslash}{\kern 0.2 em k\kern -0.45em /}
\newcommand{\lslash}{\kern 0.2 em l\kern -0.50em /}
\newcommand{\pslash}{\kern 0.2 em p\kern -0.50em /}
\newcommand{\Sslash}{\kern 0.2 em S\kern -0.50em /}
\newcommand{\Pslash}{\kern 0.2 em P\kern -0.50em /}
\newcommand{\Dslash}{\kern 0.2 em D\kern -0.65em /\kern 0.15em}
\newcommand{\slim}{\mskip 1.5mu}

\begin{document}

\title{Quark helicity distributions in transverse momentum space and transverse coordinate space}

\newcommand*{\SEU}{Department of Physics, Southeast University, Nanjing
211189, China}\affiliation{\SEU}
\newcommand*{\PKU}{School of Physics and State Key Laboratory of Nuclear Physics and
Technology, \\Peking University, Beijing 100871,
China}\affiliation{\PKU}
\newcommand*{\CHEP}{Center for High Energy Physics, Peking University, Beijing 100871, China}\affiliation{\CHEP}

\author{Zhun Lu}\email{zhunlu@seu.edu.cn}
\affiliation{\SEU}
\author{Bo-Qiang Ma}\email{mabq@pku.edu.cn}\affiliation{\PKU}\affiliation{\CHEP}

\begin{abstract}
The transverse momentum dependent helicity distributions of valence
quarks are calculated in the light-cone diquark model by adopting
two different approaches. We use the model results to analyze the
$P_{h\perp}$-dependent double spin asymmetries for $\pi^+$, $\pi^-$, and $\pi^0$ productions in semi-inclusive deep inelastic scattering,
and find that the asymmetries agree with the CLAS data in one of the
approaches. By taking the Fourier transform of the transverse
momentum dependent helicity distributions, we obtain the helicity
distributions of valence quarks in the transverse coordinate space,
and then apply them further to predict the Bessel-weighted double
spin asymmetries of $\pi^+$, $\pi^-$, and $\pi^0$ productions in
semi-inclusive deep inelastic scattering at CLAS, COMPASS, and HERMES
for the first time. The shape of the Bessel-weighted double spin
asymmetry thereby provides a direct probe on the transverse
structure of longitudinally polarized quarks.
\end{abstract}


\maketitle

\section{Introduction}

The quark helicity distribution, which measures the net helicity of
quarks in a longitudinally polarized nucleon,
\begin{align}
g_1^q(x) = f_{\,q^\uparrow/N^\uparrow}(x) - f_{\,q^\downarrow/N^\uparrow} (x)
\end{align}
is an essential observable encoding the spin structure of the nucleon~\cite{Filippone:2001ux,Aidala:2012mv}.
During the past two decades, the $x$ dependence of the helicity distribution has been studied extensively, including the parametrizations~
\cite{Gehrmann:1995ag,grsv01,Hirai:2006sr,deFlorian:2008mr,Leader:2010rb} from experimental data and model calculations~\cite{Brodsky:1994kg,Ma96,Bourrely:2001du}.
However, the transverse momentum dependence~\cite{Anselmino:2006yc,Avakian:2007xa} of the helicity distribution still remains poorly known.
In recent years, it is found that the quark transverse motion plays a crucial role~\cite{Sivers:1991prd,Collins:1993npb,Anselmino:1998plb,Boer:1998prd,
Brodsky:2002plb,Collins:2002plb,jy02} through spin-orbit correlation in certain spin phenomena, such as the single spin asymmetries measured in semi-inclusive deep inelastic scattering (SIDIS)~\cite{hermes09,hermes10,
compass06,compass10,Adolph:2012sn,Adolph:2012sp,Avakian:2010ae,Qian:2011py} and elsewhere (see Ref.~\cite{D'Alesio:2007jt} and references therein).
On the other hand, the inclusion of quark intrinsic transverse momentum can provide much richer information about the nucleon structure~\cite{Barone:2010ef,Boer:2011fh} that cannot be obtained from the collinear picture.
Thus it is necessary to revisit the quark transverse motion effect to the quark helicity distribution, especially its transverse momentum dependence that can help to build a three-dimensional partonic structure of the longitudinal polarized nucleon.

As a step toward a better understanding of this issue, we study in
this paper the transverse momentum dependent (TMD) helicity
distribution of the valence quarks, denoted by $g_{1L}^q(x,\bm
p_T^2)$, using the light-cone diquark model, which has been applied
to calculate the $x$ dependence of the helicity
distribution~\cite{Ma96}, the transversity
distribution~\cite{Ma:1997gy,Ma:2000ip} and a number of other
TMD distributions~\cite{Lu:2004au,She:2009jq,Zhu:2011zza,Zhu:2011ir}.
We will
calculate the TMD helicity distribution in two different approaches
corresponding to two different options of the unpolarized TMD
distribution. The TMD helicity distribution enters the description
of the SIDIS process when the transverse momentum of the final hadron
$P_{h\perp}$ is measured. Using the model results from the
light-cone diquark model, we calculate the double spin asymmetry
(DSA) of pion production in the SIDIS process
as a function of $P_{h\perp}$. The CLAS Collaboration
has measured~\cite{Avakian:2010ae} such asymmetry using a polarized
electron beam to collide a longitudinally polarized proton target
recently, and such measurement provides the first experimental
glimpse of the TMD helicity distribution.

Although it is possible to access the TMD helicity distribution from SIDIS data, usually one has to model the quark transverse momentum dependence in phenomenological analysis because calculating the cross section differentiated by $P_{h\perp}$ involves complicated convolution of transverse momenta~\cite{Mulders:1995dh}.
Conventionally, the convolution can be resolved through the $P_{h\perp}$-weighted procedure~\cite{Boer:1998prd,Kotzinian:1997wt} in a model-independent way, and the weighted cross section involves the transverse moments of the TMD distributions.
However, much of the information about the explicit transverse momentum dependence is lost in this way.
An alternative approach has been developed using the Bessel functions as  weighting functions in Ref.~\cite{Boer:2011xd}, where the authors show that
the convolution becomes a product of TMD distributions and TMD fragmentation functions (FFs) in the Fourier space (also referred to as the transverse coordinate space, denoted by $\bm b_T$) conjugate to $\bm p_T$.
The TMD distributions in the $\bm b_T$ space thus can be measured directly by the Bessel-weighted asymmetries whose expressions in leading twist have been given in Ref.~\cite{Boer:2011xd}.
This motivates us to examine the Bessel-weighted DSA in SIDIS by model calculation since it might be measured easier than other Bessel-weighted asymmetries.
To this end we calculate the Fourier-transformed TMD helicity distribution in the light-cone diquark model and make the prediction for the Bessel-weighted DSA of pion production in SIDIS for the first time.
The results are presented at the kinematics of CLAS, HERMES~\cite{Airapetian:2004zf}, and COMPASS~\cite{Alekseev:2010ub}, where there are ongoing SIDIS programs or there are data that could be analyzed.

The paper is organized as follows. In Sec. II, we briefly
introduce the calculation of the TMD helicity distributions of
valence quarks in the light-cone diquark model and discuss their
properties. We then calculate the DSAs of $\pi^+$, $\pi^-$, and $\pi^0$ productions in SIDIS. In Sec. III, we calculate the
valence helicity distributions in the Fourier space and present the
prediction of the Bessel-weighted DSAs of three pions in SIDIS at
CLAS, HERMES and COMPASS. Finally, we provide summaries in Sec.
IV.

\section{TMD helicity distributions in $P_{h\perp}$-dependent double spin asymmetries}

The TMD helicity distribution $g_{1L}(x,p_T^2)$ is defined through the matrix elements of the bilocal operator
\begin{align}
g_{1L}(x,\bm p_T^2) & = {1\over 2}\int {d \xi^- d^2\bm{\xi}_T \over (2\pi)^3} e^{ip\cdot \xi}\langle PS_L|\bar{\psi}(0) \nonumber\\
&\times \gamma^+\gamma_5\mathcal{L}(0,\xi)\psi(\xi) | P S_L\rangle\bigg{|}_{\xi^+=0},\label{g1ldef}
\end{align}
where $\mathcal{L}(0,\xi)$ is the gauge link ensuring the gauge invariance
of the definition. In this calculation the gauge link is set to be trivial since the helicity distribution is $T$ even.

One can insert a complete set of intermediate states $\{n\}$ into Eq.(\ref{g1ldef}):
\begin{align}
g_{1L}(x,\bm p_T^2) = & {1\over \sqrt{2}}\sum_n\int {d^3\bm P_n\over (2\pi)^32E_n}\delta\left((1-x)P^+-P_n^+\right) \nonumber\\
&\times \delta^2(\bm p_T + \bm P_{nT}) \left\{\langle PS_L|\mathcal{P}_+\psi_{(+)}(0) |n\rangle^2\right.\nonumber
\\
&\left.-\langle PS_L|\mathcal{P}_-\psi_{(+)}(0) |n\rangle^2 \right\},
\label{intepol}
\end{align}
where $\psi_{(+)} ={1\over 2}\gamma^-\gamma^+\psi$ is the ``good" component of the quark field, and $\mathcal{P}_{\pm}={1\over 2}({1\pm \gamma_5})$ are the helicity projection operators. Similarly the TMD unpolarized distribution has the form
\begin{align}
f_{1}(x,\bm p_T^2) = & {1\over \sqrt{2}}\sum_n\int {d^3\bm P_n\over (2\pi^3)2E_n}\delta\left((1-x)P^+-P_n^+\right) \nonumber\\
&\times \delta^2(\bm p_T + \bm P_{nT}) \left\{\langle P|\mathcal{P}_+\psi_{(+)}(0) |n\rangle^2\right.\nonumber
\\
&\left.+\langle P|\mathcal{P}_-\psi_{(+)}(0) |n\rangle^2 \right\}.
\label{inteun}
\end{align}

The TMD helicity distribution has been calculated by different
models, such as the diquark spectator  model~\cite{Jakob:1997npa,Bacchetta:2008prd}, the light-cone constituent quark model~\cite{Pasquini:2008ax}, the covariant parton model~\cite{Efremov:2009ze}, and the bag model~\cite{Avakian:2010};
recently it has also been studied by lattice QCD~\cite{Musch:2010ka}.
Here, we adopt the light-cone
diquark model~\cite{Ma96,Brodsky01} to calculate the valence TMD
helicity distributions.
The above mentioned models have been widely applied to calculate various TMD distributions~\cite{Lu:2004au,She:2009jq,Zhu:2011zza,Zhu:2011ir,
Jakob:1997npa,Bacchetta:2008prd,Pasquini:2008ax,
Efremov:2009ze,Avakian:2010,Gamberg:2007wm}, including the $T$-even and $T$-odd distributions.
Here a few comments on these models are in order.
The common feature of the diquark spectator model~\cite{Jakob:1997npa,Bacchetta:2008prd} and the light-cone diquark model~\cite{Ma96,Brodsky01} is that when every valence quark is probed, one treats the other part of the nucleon as a spectator and uses the diquark quantum numbers to effectively take into account the spectator contributions.
The advantage of the diquark-type models is that the complicated many-particle
system can be treated by a two-particle technique.
In the light-cone constituent quark model~\cite{Pasquini:2008ax}, the proton state is described in terms of three free on-shell valence quarks.
The TMD distributions are calculated through the light-cone wavefunctions that take into account the full dynamics of three valence quarks.
In the covariant parton model approach~\cite{Efremov:2009ze}, the partons are free, and are assumed to be
described in terms of three-dimensional spherically symmetric momentum distributions in the nucleon rest frame. Different from the above-mentioned models, the bag model is the quark model which incorporates confinement, i.e., it describes the nucleon as three massless quarks confined inside a sphere, modeled by the bag boundary condition~\cite{Avakian:2010}. Both the covariant parton model and the bag model support the Gaussian dependence of the quark transverse momenta~\cite{Schweitzer:2010}, while in the diquark model and the light-cone constituent model the transverse momentum dependent can be non-Gaussian.

In the SU(6) quark-diquark model, the proton
state with a spin component $S_z= +\frac{1}{2}$ in the instant form
can be written as
\begin{eqnarray}
|P\uparrow\rangle&=&\frac{1}{3}\sin\theta\varphi_V\left[(ud)^0u^\uparrow
- \sqrt{2}(ud)^1u^\downarrow - \sqrt{2}(uu)^0d^\uparrow \right.\nonumber \\
&&\left.+
2(uu)^1d^\downarrow\right]+\cos\theta\varphi_S(ud)^Su^\uparrow,~~~~\label{pup}
\end{eqnarray}
where the two quarks inside the brackets form the scalar ($S$) or the vector ($V$) diquarks, $\uparrow$/$\downarrow$ denotes the positive/negative helicity of quarks,
$\theta$ is the mixing angle that breaks the SU(6) symmetry when $\theta\neq\pi/4$,
and $\varphi_D(x,\bm p_T^2)$ is the wave function in the momentum space for the quark-diquark system, which satisfies the normalization condition $\int dx d^2\bm p_T\left|\varphi_D(x,\bm p_T^2)\right|^2=3$.
So the correct valence quark numbers are satisfied. In addition, in the light-cone diquark model the total valence quark momentum fractions can be adjusted from the phenomenological consideration.

Notice that the proton wavefunction in Eq.~(\ref{pup}) is in the
instant form dynamics with equal $t$ in an ordinary frame; however,
the parton distributions are usually expressed in the infinite
momentum frame or in the light-front dynamics on the light cone.
Thus, we need to transform the wavefunction in the instant form to
that in the light-cone form. The connection between the two forms is
through a Melosh-Wigner rotation~\cite{Melosh74,Ma91}. For a
spin-${1}/{2}$ particle, we use $q_T$ and $q_F$ to denote the
instant and light-cone spinors, respectively, and the Melosh-Wigner
rotation is known to be~\cite{Ma91}
\begin{eqnarray}
\left(\begin{array}{c}
q_T^\uparrow\\
q_T^\downarrow
\end{array}\right)
&=&\omega\left(\begin{array}{cc}
p^++m & -p^R\\
p^L & p^++m
\end{array}\right)\left(\begin{array}{c}
q_F^\uparrow\\
q_F^\downarrow
\end{array}\right)\nonumber\\
&\equiv& \bm{M}^{1/2}\left(\begin{array}{c}
q_F^\uparrow\\
q_F^\downarrow
\end{array}\right),
\label{Melosh}
\end{eqnarray}
where $p^{R,L}=p^1\pm i p^2$, $p^+ = x\mathcal{M}_D$,
$\omega=\left[(x\mathcal{M}_D+m)^2+\bm p^2_T\right]^{-1/2}$ with
$\mathcal{M}_D^2=\frac{m^2+\bm p^2_T}{x}+\frac{m_D^2+\bm p^2_{T}}{1-x}$,
$m$ and $m_D$ are mass parameters for the quark and diquark,  and $\bm{M}^{1/2}$ denotes the Melosh-Wigner rotation matrix for the spin-$1/2$ particles.
For the spin-0 scalar diquark, there is no such  transformation.
For the spin-1 vector diquarks, the transformation is represented by a $3\times3$
matrix $\bm{M}^1$, whose explicit expression can also be found in
Refs.~\cite{Ahluwalia:1993xa,Ma2002}.
In practice, we will not use the explicit form, for it does not appear in the final expression for the spectator debris due to the unitary property $\bm{M}^{1\dag}\bm{M}^1=I$.

Substituting Eq.~(\ref{Melosh}) into (\ref{pup}) and using Eq.~(\ref{intepol}), we obtain the TMD helicity distributions for the valence $u$ and $d$ quarks
\begin{align}
g_{1L}^{uv}(x,\bm{p}_T^2) =&\frac{1}{16\pi^3}\cos^2\theta\varphi_S^2W_S(x,\bm{p}_T)\nonumber\\
&-\frac{1}{48\pi^3}\times\frac{1}{3}
\sin^2\theta\varphi_V^2W_V(x,\bm{p}_T),\label{g1luv2}\\
g_{1L}^{dv} (x,\bm{p}_T^2) = &-\frac{1}{24\pi^3}\times\frac{1}{3}
\sin^2\theta\varphi_V^2W_V(x,\bm{p}_T),\label{g1ldv2}
\end{align}
where $W_D$ is the Melosh-Wigner rotation factor for the helicity distribution with the form~\cite{Ma96,Ma91}
\begin{eqnarray}
W_D (x, \bm p_T) = \frac{(x\mathcal{M}_D+m)^2-\bm p^2_T}{(x\mathcal{M}_D+m)^2+\bm p^2_T}. \label{rotation}
\end{eqnarray}
Similarly the TMD unpolarized distributions
are calculated as
\begin{eqnarray}
\label{unp}
f_1^{uv}(x,\bm{p}_T)&=&\frac{1}{16\pi^3}\times(\frac{1}{3}
\sin^2\theta\varphi_V^2
+\cos^2\theta\varphi_S^2),\label{f1uv}\\
f_1^{dv}(x,\bm{p}_T)&=&\frac{1}{8\pi^3}\times\frac{1}{3}
\sin^2\theta\varphi_V^2.\label{f1dv}
\end{eqnarray}
The TMD helicity distributions thus are obtained by combining Eqs.
(\ref{g1luv2}), (\ref{g1ldv2}) (\ref{f1uv}) and (\ref{f1dv}),
\begin{align}
g_{1L}^{uv}(x,\bm{p}_T^2) =&\left[f_1^{(uv)}(x,\bm{p}_T^2)
-\frac{1}{2}f_1^{(dv)}(x,\bm{p}_T^2)\right]W_S(x,\bm{p}_T)\nonumber\\
&-\frac{1}{6}f_1^{(dv)}(x,\bm{p}_T^2)W_V(x,\bm{p}_T),\label{g1luv}\\
g_{1L}^{dv} (x,\bm{p}_T^2) = &-\frac{1}{3}f_1^{(dv)}(x,\bm{p}_T^2)W_V(x,\bm{p}_T). \label{g1ldv}
\end{align}
Performing the integration over $\bm p_T$, one arrives at the integrated helicity distributions
\begin{align}
g_{1}^{uv}(x)& = \int d^2\bm p_T g_{1L}^{uv}(x,\bm{p}_T^2),\\
g_{1}^{dv}(x) &= \int d^2\bm p_T g_{1L}^{dv}(x,\bm{p}_T^2),
\end{align}
which have already been calculated in Ref.~\cite{Ma96} in
the light-cone diquark model.

Equations.~(\ref{g1luv}) and (\ref{g1ldv}) show that the helicity distributions are connected to the unpolarized distributions in the light-cone diquark model.
Here, we employ two approaches to calculate $g_{1L}(x,\bm p_T^2)$, corresponding to two different choices for $f_1^q(x,\bm p_T^2)$.
In Approach 1 we adopt the Brodsky-Huang-Lepage
prescription~\cite{Brodsky82,Huang1994} for the wavefunction in the momentum space:
\begin{eqnarray}
\varphi_D=A_D\exp\left\{-\frac{1}{8\alpha_D^2}
[\frac{m_q^2+{\bm p}_T^2}{x}+\frac{m_D^2+{\bm p}_T^2}{1-x}]\right\}, \label{bhl}
\end{eqnarray}
to calculate the unpolarized distributions $f_1^q(x,\bm p_T^2)$.
In Approach 2 we factorize the $x$ dependence and the $\bm p_T$ dependence of $f_1^q(x,\bm p_T^2)$ and choose the Gaussian form for the latter one in further
\begin{align}
f_1^q(x,\bm p_T^2) = f_1^q(x){1\over \pi \langle p_T^2 \rangle_f^q} \exp\left(-\bm p_T^2 \over \langle p_T^2 \rangle_f^q\right), \label{af2}
\end{align}
which has been used in many phenomenological applications. For the integrated distribution $f_1^q(x)$ appearing in Eq.~(\ref{af2}), we use the phenomenological input, i.e., the leading order set of the MSTW2008 parametrization~\cite{Martin2009}.
In both approaches the $p_T$-dependence for the polarized quarks is different from the Gaussian distribution which is adopted in
Ref.~\cite{Anselmino:2006yc}.

\begin{figure}
\begin{center}
\includegraphics[width=0.7\columnwidth]{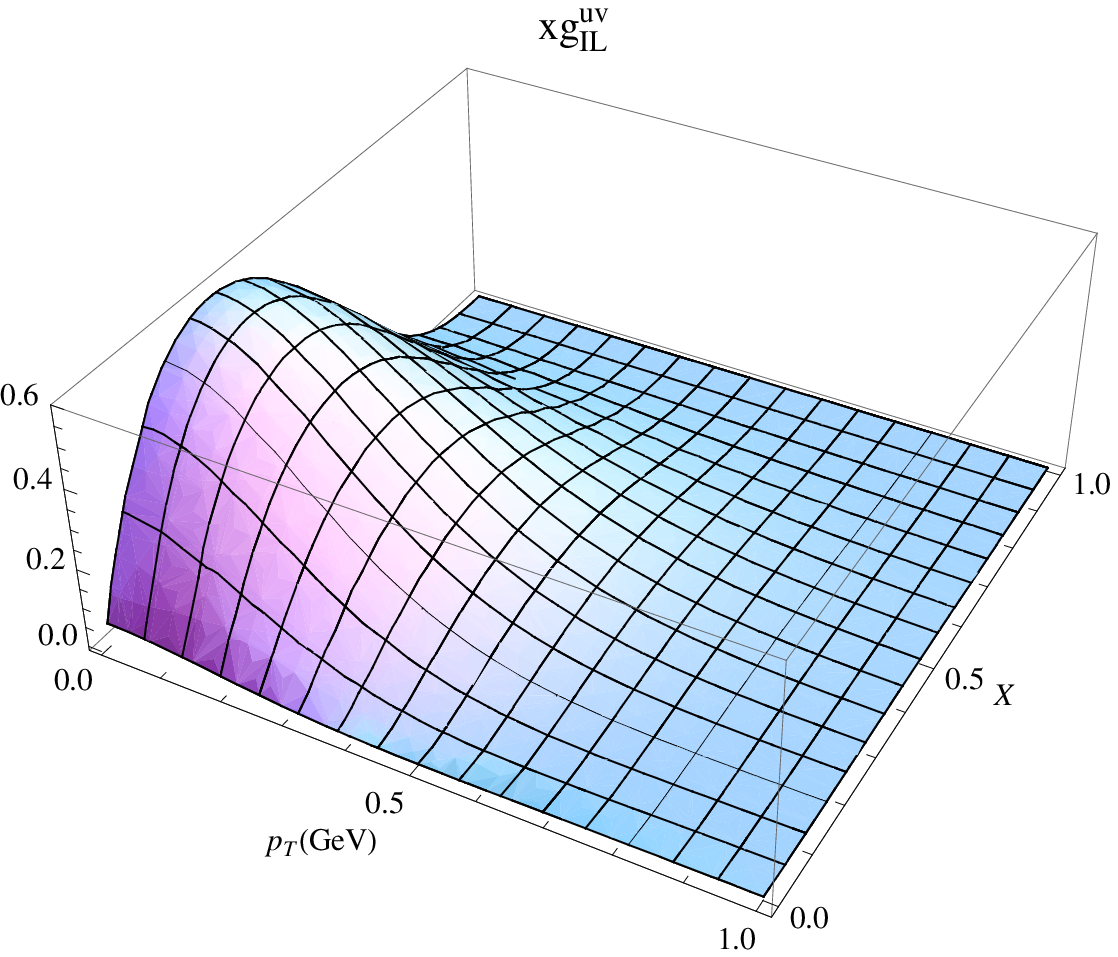}\\
  \includegraphics[width=0.7\columnwidth]{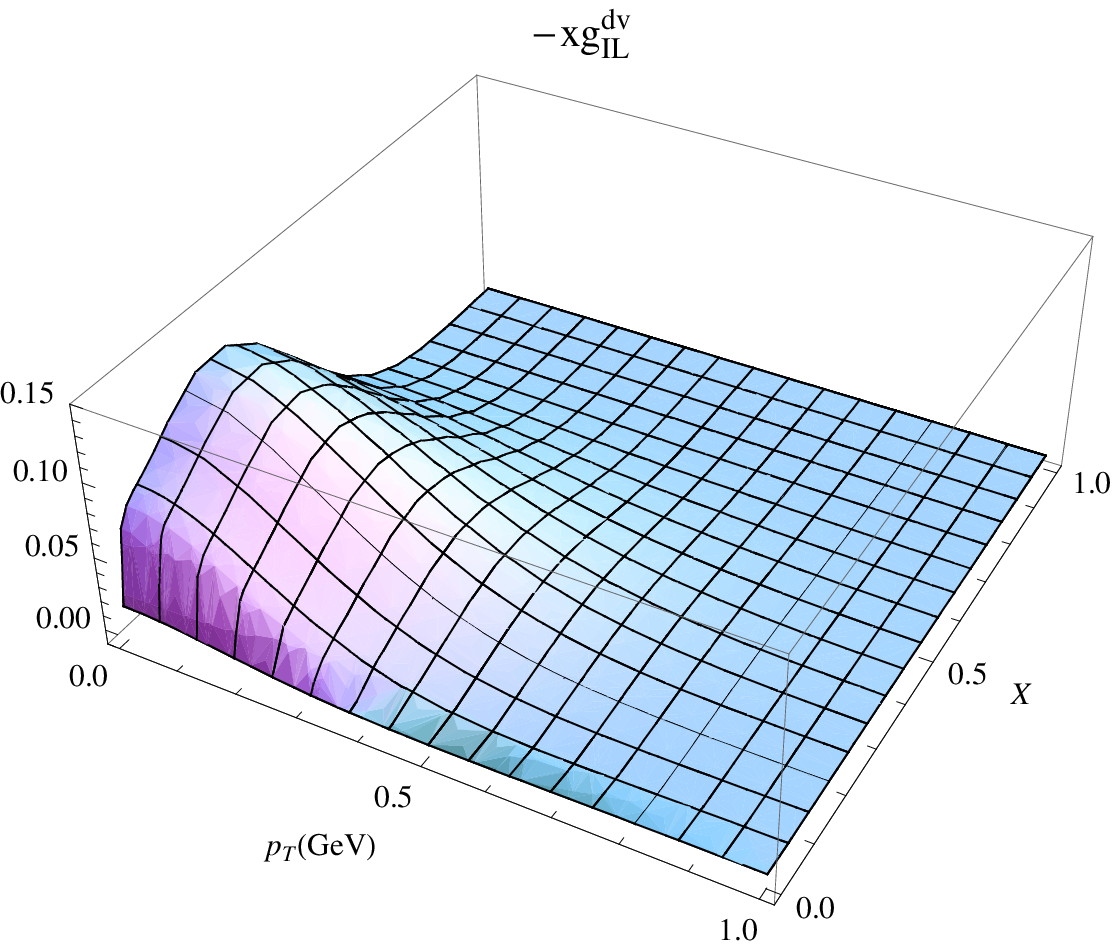}
  \caption{The three-dimensional demonstration of $xg_{1L}^{uv}(x,\bm p_T^2)$ (upper panel) and $-g_{1L}^{dv}(x,\bm p_T^2)$ (lower panel) calculated by the light-cone diquark model in Approach 2.}\label{g1L3d}
\end{center}
\end{figure}

As a demonstration, in the upper panel and the lower panel of Fig.~\ref{g1L3d}, we show the three-dimensional shapes of $xg_{1L}^{uv}(x,\bm p_T^2)$ and $-xg_{1L}^{dv}(x,\bm p_T^2)$ calculated from the light-cone diquark model using Approach 2, respectively.
In this calculation the unpolarized valence quark PDFs $f_1^{uv}(x)$ and $f_1^{dv}(x)$ are chosen at the scale $Q=1.2~\text{GeV}$, and the Gaussian width for the unpolarized  $u$ and $d$ quarks  is adopted as $\langle p_T^2\rangle_f^q =0.25\,\text{GeV}$~\cite{Anselmino:2005nn}.
Also we choose the mass parameters as $m_q=0.35\, \text{GeV}$, $M_S=0.6\,\text{GeV}$, and $M_V=0.8\,\text{GeV}$, so that the sum of the quark and the diquark masses is larger than the proton mass in order to form a stable bound state.

Keeping the transverse momentum unintegrated, we can also explore the properties of the TMD distributions of quarks with positive helicity [denoted by $q^{\uparrow}(x,\bm p_T^2)$] and those with negative helicity [denoted by $q^{\downarrow}(x,\bm p_T^2)$], which can be obtained from the combination
of $f_1^q(x,\bm p_T^2)$ and $g_{1L}^q(x,\bm p_T^2)$,
\begin{align}
q^{\uparrow}(x,\bm p_T^2) &={1\over 2} \left(f_1^{q}(x,\bm p_T^2)+g_{1L}^{q}(x,\bm p_T^2)\right), \\
q^{\downarrow}(x,\bm p_T^2) &={1\over 2} \left(f_1^{q}(x,\bm p_T^2)-g_{1L}^{q}(x,\bm p_T^2)\right).
\end{align}

\begin{figure}
\begin{center}
\includegraphics[width=\columnwidth]{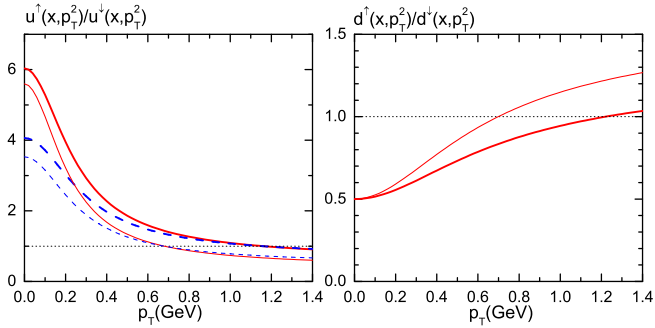}
  \caption{The $p_T$ profiles of $u_v^\uparrow(x,\bm p_T^2)/u_v^\downarrow(x,\bm p_T^2)$ (left panel) and $d_v^\uparrow(x,\bm p_T^2)/d_v^\downarrow(x,\bm p_T^2)$ at $x=0.3$ (thick lines) and $x=0.15$ (thin lines) in Approach 1 (solid line)and Approach 2 (dashed line).}\label{udratio}
\end{center}
\end{figure}

In Fig.~\ref{udratio}, we plot the ratios $q^\uparrow(x,\bm
p_T^2)/q^\downarrow(x,\bm p_T^2)$ for $q = u_v$ and $d_v$ as
functions of $p_T$ at $x=0.3$ (thick lines) and $x=0.15$ (thin
lines) in the light-cone diquark model using Approach 1 and Approach
2. From our model calculation we see that at small $p_T$, the
density of the valence $u$ quark with positive helicity is several times
larger than that with negative helicity at small $p_T$; but the
difference between $u_v^\uparrow(x,\bm p_T^2)$ and
$u_v^\downarrow(x,\bm p_T^2)$ reduces with increasing $p_T$. For the
$d$ valence quark, the ratios of the positive helicity density and
negative helicity density are the same in both approaches. At
smaller $x$ and larger transverse momentum, the $u$ quark has net
negative helicity, while the $d$ quark tends to have net positive
helicity. These features are originated from the quark transverse motion,
which can be seen from the Melosh-Wigner rotation factor
in Eq.~(\ref{rotation}) that is negative for large $\bm p_T$ and
small $x$.

In principle, the TMD helicity distributions can be accessed in the SIDIS
process~\cite{Mulders1996} by employing the longitudinally polarized
lepton beam off the longitudinally polarized nucleon target, when
the transverse momentum of the final fragmenting hadron is measured.
The CLAS Collaboration at JLab has performed such a measurement for
the $\pi^+$, $\pi^-$, and $\pi^0$ productions~\cite{Avakian:2010ae}.
In the following we will use our model resulting TMD helicity
distributions to study the phenomenology of pion production in the
double polarized SIDIS process.

The differential cross section of the double longitudinally polarized SIDIS has the general form~\cite{Diehl:2005pc,Bacchetta:2006tn}
\begin{align}
\frac{d\sigma}{dx dy\,dz dP^2_{h\perp} d\phi} = \,& \frac{2\pi \alpha^2}{x y Q^2}
 \Bigl( 1+ \frac{\gamma^2}{2x} \Bigr) \left\{ A(\varepsilon)F_{UU,T}\right.\nonumber\\
 &\left. +\, \lambda_e S_{||} B(\varepsilon)  \,\,F_{LL}\right\}, \label{sig_ll}
\end{align}
where the higher-twist contribution~\cite{Anselmino:2006yc} has not been considered.
The factors $A(\varepsilon)$ and $B(\varepsilon)$ are the depolarization factors with the forms
\begin{align}
A(\varepsilon)&=\frac{y^2}{2(1-\varepsilon)} \approx \left(1-y +\tfrac{1}{2}\slim y^2\right), \\
B(\varepsilon)&=\frac{y^2}{2(1-\varepsilon)}\sqrt{1-\varepsilon^2}
\approx y\left(1 -\tfrac{1}{2}\slim y\right).
\end{align}
The approximations hold in the limit $\gamma=2Mx/Q\rightarrow 0$.
In the leading order of TMD factorization~\cite{Ji:2004wu} the two structure functions $F_{UU,T}$ and $F_{LL}$ are expressed as
\begin{align}
F_{UU,T} =&\, x\sum_q e_q^2\int d^2\bm p_T \int d^2 \bm K_T \delta^2(z\bm p_T-\bm P_{h\perp}+\bm K_T)\nonumber\\
 &\times f_1^q(x, \bm p_T^2) D_1^q(z,\bm K_T^2), \label{FUU}\\
F_{LL}  = &\,x\sum_q e_q^2\int d^2\bm p_T \int d^2 \bm K_T \delta^2(z\bm p_T-\bm P_{h\perp}+\bm K_T)\nonumber\\
&\times g_{1L}^q(x, \bm p_T^2) D_1^q(z,\bm K_T^2).\label{FLL}
\end{align}
Here, $\bm K_T$ is the transverse momentum of the final pion with respect to the momentum of the fragmenting quark, and $\bm P_{h\perp}$ is the transverse momentum of the pion with respect
to the photon momentum in the target rest frame.
For the TMD FFs needed in the calculation,
we assume that their $K_T$ dependence has a Gaussian form
\begin{align}
D_1^q\left(z,\bm K_T^2\right)=D_1^q(z)\, \frac{1}{\pi \langle K_T^2\rangle}
\, \exp(-\bm K_T^2/\langle K_T^2\rangle)\,,\label{tmdd1}
\end{align}
and we choose the Gaussian width $\langle K_T^2\rangle = 0.20\,\text{GeV}$~\cite{Anselmino:2005nn} for the $u$ and $d$ quarks. For the collinear FFs we adopt the leading-order set of the Kretzer parametrization~\cite{kretzer2000}.

The DSA in SIDIS can be measured through
\begin{align}
A_{LL} (x,z,P_T) &= {d\sigma^{\uparrow \uparrow} - d\sigma^{\uparrow \downarrow}
-d\sigma^{\downarrow \uparrow}+ d\sigma^{\downarrow \downarrow}
\over d\sigma^{\uparrow \uparrow}+ d\sigma^{\uparrow \downarrow}
+d\sigma^{\downarrow \uparrow}+ d\sigma^{\downarrow \downarrow}} \nonumber\\
&= { \frac{1}{xyQ^2}
 \Bigl( 1+ \frac{\gamma^2}{2x} \Bigr) B(\varepsilon) F_{LL}\over
 \frac{1}{xyQ^2} \Bigl( 1+ \frac{\gamma^2}{2x} \Bigr) A(\varepsilon) F_{UU,T}},
\end{align}
where $\uparrow$/$\downarrow$ denotes the positive/negative helicity of the lepton and the nucleon, respectively.
On the other hand, one may define the DSA that is scaled by the depolarization factors
\begin{align}
A_1 = {A(\varepsilon)\over B(\varepsilon)}A_{LL}
 &= { \frac{1}{xyQ^2}
 \Bigl( 1+ \frac{\gamma^2}{2x} \Bigr) F_{LL}\over
 \frac{1}{xyQ^2} \Bigl( 1+ \frac{\gamma^2}{2x} \Bigr) F_{UU,T}}, \label{A1}
\end{align}
which is consistent with the measurement by the CLAS Collaboration~\cite{Avakian:2010ae}.

\begin{figure}
\begin{center}
  \includegraphics[width=\columnwidth]{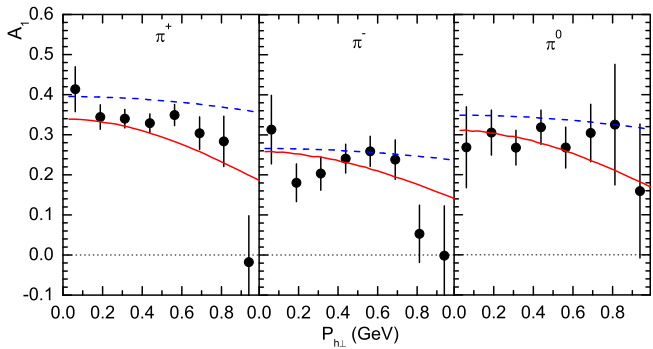}
  \caption{Double spin asymmetries $A_1$ for the $\pi^+$, $\pi^-$, and $\pi^0$ productions as functions of the pion transverse momentum $P_{h\perp}$. The dashed and the solid curves correspond to the predictions using
the TMD unpolarized and helicity distributions from Approach
1 and 2 of the light-cone diquark model, respectively. Data are from \cite{Avakian:2010ae}. }\label{A1clas}
\end{center}
\end{figure}

In Fig.~\ref{A1clas}, we plot the $P_{h\perp}$-dependent DSAs defined in Eq.~(\ref{A1}) for the $\pi^+$, $\pi^-$, and $\pi^0$ productions at the CLAS kinematics,
\begin{align}
&E_e=5.7\, \text{GeV},~~ 0.12<x<0.48,~~
y<0.85,  \nonumber\\
&0.4<z<0.7,~~ 0.9 \, \text{GeV}^2<Q^2 <5.4\, \textrm{GeV}^2,\label{kine-clas}\\
&W^2>4\,\textrm{GeV}^2, ~~ M_x(e \pi^0) > 1.4 \,\textrm{GeV}.\nonumber
\end{align}
and compare them with the CLAS data~\cite{Avakian:2010ae}.
The dashed and the solid curves correspond to the predictions using
the TMD unpolarized and helicity distributions from Approach
1 and 2 of the light-cone diquark model, respectively.
It is found that the TMD distributions in Approach 2 agree with the CLAS data generally, that is, we predict that the asymmetries for all three pion channels have the tendency to fall off at large $P_{h\perp}$.
However, the model has the difficulty to explain the positive slope for $\pi^-$ production at the moderate $P_T$ region.
In Ref.~\cite{Avakian:2010ae} the $P_T$ dependence of the double-spin asymmetry was interpreted by difference widths of the quark TMD distributions with
different flavor and polarizations~\cite{Anselmino:2006yc} resulting from different orbital motion of quarks polarized in the direction of the proton spin and opposite to it~\cite{Brodsky:1994kg,Avakian:2007xa}.
In Table.~\ref{tab1} we list the widths for the polarized quarks that
are defined as
\begin{align}
\langle p_T^2 \rangle_g^q =
{\int_{x_{\text min}}^{x_{\text max}} dx \int d^2\bm p_T \, p_T^2\, g_{1L}^q(x,\bm p_T^2)
\over \int_{x_{\text min}}^{x_{\text max}} dx
\int d^2\bm p_T g_{1L}^q(x,\bm p_T^2)},
\end{align}
in the light-cone diquark, together with the widths for the unpolarized quarks.
The integrating limits for $x$ are chosen as $x_{\text min}=0.1$ and $x_{\text max}=0.9$, corresponding to the valence region.
In both cases, the widths for the polarized $u$ and $d$ valence quarks are almost the same.
The ratio $\langle p_T^2 \rangle_g / \langle p_T^2 \rangle_f $ is around $0.52-0.56$ in Approach 2, while that in Approach 1 is $0.8$.

\begin{table}
\begin{tabular}{|c|c|c|c|c|}
  \hline
   & $\langle p_T^2\rangle_f^{uv}$& $\langle p_T^2\rangle_f^{dv}$  & $\langle p_T^2\rangle_g^{uv}$& $\langle p_T^2\rangle_g^{dv}$\\
   \hline
   \hline
  Approach 1 & 0.25\,GeV$^2$  & 0.25\,GeV$^2 $ &0.14\,GeV$^2$  &0.13\,GeV$^2$ \\
  \hline
  Approach 2 & 0.10\,GeV$^2$ & 0.10 \,GeV$^2$ & 0.08\,GeV$^2$&0.08\,GeV$^2$ \\
  \hline
\end{tabular}
\caption{The mean square momenta of the unpolarized and the polarized quarks in Approach 1 and Approach 2.}\label{tab1}
\end{table}

\section{Helicity distribution in Fourier space and Bessel-weighted DSA}

Because the structure functions given in Eqs.(\ref{FUU}) and (\ref{FLL}) involve the convolution of the transverse momenta of partons, it is not
easy to obtain directly the information of TMD distributions when $P_{h\perp}$ is measured.
In Ref.~\cite{Boer:2011xd}, Boer, Gamberg, Musch and Prokudin developed a framework to avoid the convolution of the transverse momenta
through the so-called Bessel-weighted procedure.
In this framework, instead of the TMD distributions, the Fourier transformation of them [denoted by $\tilde f^q(x,\bm b_T^2)$] can be accessed directly by measuring the Bessel-weighted asymmetries, where $\bm b_T$ is the transverse position of partons conjugate to $\bm p_T$.
Thus, $\tilde f^q(x,\bm b_T^2)$ are also referred to as the parton distributions in the transverse position space.
Once $\tilde f^q(x,\bm b_T^2)$ are measured, the TMD distributions can be obtained by the inverse Fourier transformation on $\tilde f^q(x,\bm b_T^2)$.
The Fourier-Bessel transforms is also useful in studying the factorization as well as the scale dependence of the TMD cross section
~\cite{Parisi:1979se,Jones:1980my,Collins:1981uw,Ellis:1981sj,
Collins:1984kg,Ji:2004xq,Collins:2011}.

In this section, we study the helicity distribution in the transverse position space and the related phenomenology in SIDIS processes using the Bessel-weighted framework developed in Ref.~\cite{Boer:2011xd}. Using the identity
\begin{align}
\delta^2(z\bm p_T-\bm P_{h\perp}+\bm K_T) =
\int {d^2 \bm b_T\over (2\pi)^2}  e^{i\bm b_T\cdot(z\bm p_T-\bm P_{h\perp}+\bm K_T)},
\end{align}
the integrations over the transverse momenta in Eqs.~(\ref{FUU}) and (\ref{FLL}) are deconvoluted~\cite{Boer:2011xd,Ji:2004xq}:
\begin{align}
F_{UU,T} =&\, x\sum_q e_q^2\int {d|\bm b_T| \over 2\pi}|\bm b_T| J_0(|\bm b_T||\bm P_{h\perp}|)\nonumber\\
 &\times \tilde{f}_1^q(x, z^2\bm b_T^2) \,\tilde D_1^q(z,\bm b_T^2), \label{FUUbesl}\\
F_{LL}  = &\,x\sum_q e_q^2\int {d|\bm b_T| \over 2\pi}|\bm b_T| J_0(|\bm b_T||\bm P_{h\perp}|)\nonumber\\
 &\times \tilde{g}_{1L}^q(x, z^2\bm b_T^2) \,\tilde D_1^q(z,\bm b_T^2),\label{FLLbsel}
\end{align}
where the functions with a tilde are the Fourier transforms of the TMD distributions and FFs,
\begin{align}
\tilde{f}_1^q(x, z^2\bm b_T^2) &=\int d^2 \bm p_T  e^{iz\bm b_T \cdot \bm p_T }
f_1^q(x, \bm p_T^2),\\
\tilde{g}_{1L}^q(x, z^2\bm b_T^2)
 &=\int d^2 \bm p_T  e^{iz\bm b_T \cdot \bm p_T }
g_{1L}^q(x, \bm p_T^2), \\
\tilde D_1^q(z,\bm b_T^2) & = \int d^2 \bm p_T  e^{i\bm b_T \cdot \bm p_T }
D_1^q(z, \bm p_T^2).
\end{align}
The Bessel-weighted asymmetry corresponding to the DSA defined in Eq.~(\ref{A1}) can be obtained by the weighting function $J_0(|\bm {\mathcal B}_T||\bm P_{h\perp}|)$, with $\bm {\mathcal B}_T$ the transverse coordinate conjugate to $\bm P_{h\perp}$~\cite{Boer:2011xd}:
\begin{align}
&A_1^{J_0(|\bm {\mathcal B}_T||\bm P_{h\perp}|)}(\bm {\mathcal B}_T)
\nonumber\\
 &= { \frac{1}{xyQ^2}
 \Bigl( 1+ \frac{\gamma^2}{2x} \Bigr)\int d|\bm P_{h\perp}| |\bm P_{h\perp}|  J_0(|\bm {\mathcal B}_T||\bm P_{h\perp}|)  F_{LL}\over
 \frac{1}{xyQ^2} \Bigl( 1+ \frac{\gamma^2}{2x} \Bigr)\int d|\bm P_{h\perp}| |\bm P_{h\perp}|  J_0(|\bm {\mathcal B}_T||\bm P_{h\perp}|) F_{UU,T}} \nonumber \\
 &= { \frac{1}{yQ^2}
 \Bigl( 1+ \frac{\gamma^2}{2x} \Bigr)\,\sum_q e_a^2\, \tilde{g}_{1L}^q(x, z^2\bm {\mathcal B}_T^2) \,\tilde{D}_1^q(z, \bm{\mathcal B}_T^2)\over
 \frac{1}{yQ^2} \Bigl( 1+ \frac{\gamma^2}{2x} \Bigr)\,\sum_q e_a^2\, \tilde{f}_1^q(x, z^2\bm {\mathcal B}_T^2) \,\tilde{D}_1^q(z, \bm{\mathcal B}_T^2) },
 \label{A1besl}
\end{align}
where convolutions of TMD distributions and FFs become simple products.
Another advantage of the Bessel-weighted asymmetries is that the universal soft factors appearing in the numerator and the denominator of~Eq.(\ref{A1besl}) cancel~\cite{Boer:2011xd}.
In principle the hard parts $H_{LL}$ and $H_{UU,T}$ for the polarized and unpolarized subprocess~\cite{Ji:2004xq} should also present in the numerator and the denominator of Eq.~(\ref{A1besl}).
At higher order they may receive different corrections.
However, the calculation of the hard parts at one-loop order shows that $H_{LL}=H_{UU,T}$.
Thus it is a quite good approximation to use Eq.~(\ref{A1besl}) to perform numerical calculation.
Furthermore, in the case of $P_{h\perp}$-dependent asymmetry the leading-order expression (\ref{A1}) can describe the data, as we have shown in the previous section.

\begin{figure}
\begin{center}
\includegraphics[width=0.7\columnwidth]{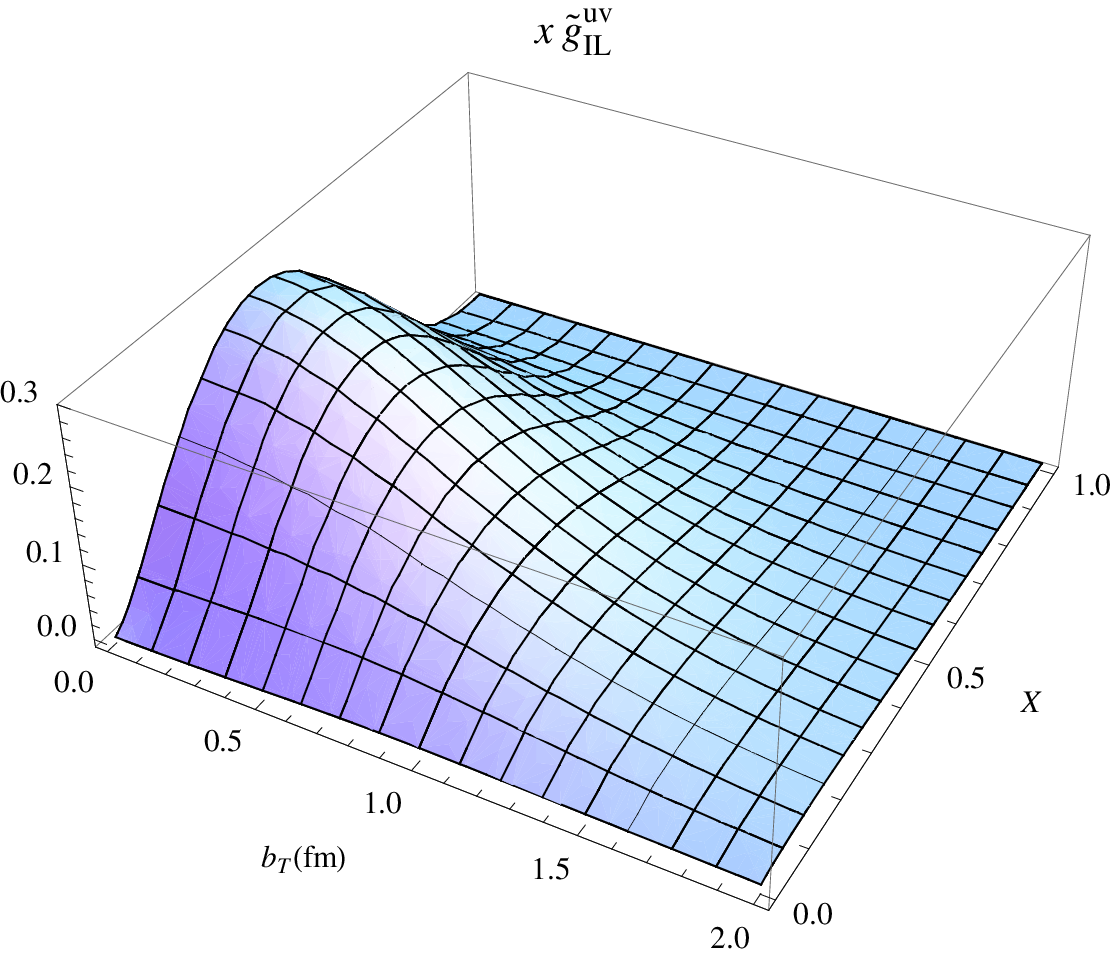}\\
  \includegraphics[width=0.7\columnwidth]{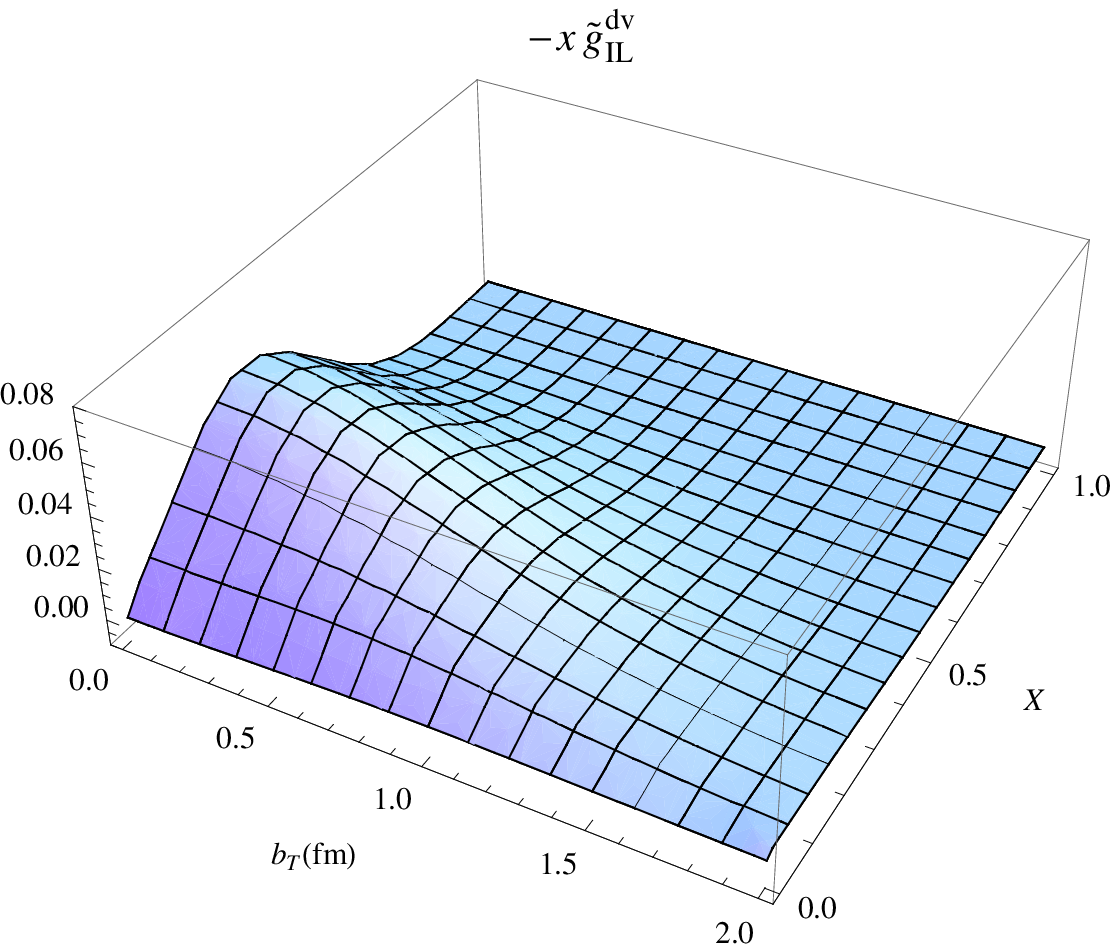}
  \caption{The three-dimensional demonstration of $x\tilde g_{1L}^{uv}(x,\bm b_T^2)$ (upper panel) and $-\tilde g_{1L}^{dv}(x,\bm b_T^2)$ (lower panel) calculated by the light-cone diquark model in Approach 2.}\label{g1lud-besl}
\end{center}
\end{figure}

We use Approach 2 in the light-cone diquark model to study the Bessel-weighted DSA.
According to Eqs.~(\ref{ap2}) and (\ref{tmdd1}), in this approach the unpolarized distribution and FFs in the Fourier space have the analytic forms
\begin{align}
\tilde{f}_1^q(x, \bm b_T^2) &=\int d^2 \bm p_T  e^{i\bm b_T \cdot \bm p_T }
{f_1^q(x)\over \pi \langle p_T^2 \rangle_f^q} \exp\left(-\bm p_T^2 \over \langle p_T^2 \rangle_f^q\right)\nonumber\\
&= f_1^q(x)\, \exp\left(-\bm b_T^2  \langle p_T^2 \rangle_f^q\over 4\right) \\
\tilde D_1^q(z,\bm b_T^2) & = \int d^2 \bm K_T  e^{i \bm b_T \cdot \bm K_T }
D_1^q(z, \bm K_T^2) \nonumber\\
& = D_1^q(z)\, \exp\left(-\bm b_T^2  \langle K_T^2 \rangle\over 4\right).
\end{align}
Thus the Gaussian widths for $\tilde{f}_1^q(x, \bm b_T^2)$ and
$\tilde D_1^q(z,\bm b_T^2)$ are $4/\langle p_T^2 \rangle_f^q$ and
$4/\langle K_T^2 \rangle$, respectively.
In our model the helicity distributions in the Fourier space cannot be obtained analytically.
We calculate them numerically instead, and show the three-dimensional shapes of $x\tilde g_{1L}^{uv}(x,\bm b_T^2)$ and $-x\tilde g_{1L}^{dv}(x,\bm b_T^2)$ in Fig.~\ref{g1lud-besl}. We see that both of them are sizable in the region $ b_T<1\, \text{fm}$, and approach zero at $ b_T=2 \,\text{fm}$.

\begin{figure}
\begin{center}
\includegraphics[width=\columnwidth]{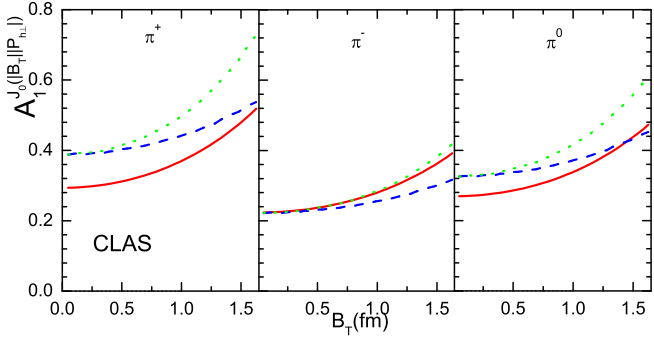}
\caption{The Bessel-weighted DSAs $A_1^{J_0(|\bm{\mathcal B}_T||\bm P_{h\perp}|)}$ for $\pi^+$, $\pi^-$, and $\pi^0$ productions as functions of $\mathcal B_T$ at CLAS.
The solid lines are from the Approach 2 of the light-con diquark model, while the dashed line and the dotted lines are from the Gaussian ansatz for the TMD helicity distributions with $\langle p_T^2\rangle_g^q = 0.17 ~\textrm{GeV}$ and
$0.10~\text{GeV}^2$, respectively.
}\label{a1clas-j0}
\end{center}
\end{figure}

The Bessel-weighted DSAs as functions of $\mathcal B_T$ for $\pi^+$, $\pi^-$, and $\pi^0$ productions at the CLAS kinematics (\ref{kine-clas}) are plotted in Fig.~\ref{a1clas-j0}.
The solid curves are the results from Approach 2 in the light-cone diquark model.
As a comparison, we also present the prediction by using the TMD helicity distributions adopted in Ref.~\cite{Anselmino:2006yc}, where $g_1^q(x,\bm p_T^2)$ has a Gaussian dependence for $p_T$,
\begin{align}
g_{1L}^q(x,\bm p_T^2) = g_1^q(x){1\over \pi \langle p_T^2 \rangle_g^q} \exp\left(-\bm p_T^2 \over \langle p_T^2 \rangle_g^q\right), \label{ap2}
\end{align}
and $g_1^q(x)$ is taken from the GRSV2000 leading-order parametrization~\cite{grsv01}.
The results are shown by the dashed and dotted curves for
$\langle p_T^2\rangle_g^q=0.17\,\text{GeV}^2$ and 0.10 GeV$^2$,
respectively.
Although the TMD helicity distributions from the light-cone diquark model (Approach 2) and from the Gaussian ansatz can both describe the CLAS data, their predictions for the Bessel-Weighted DSAs are apparently different, especially for the $\pi^+$ production.
Thus we expect that the measurement of the Bessel-weighted DSAs can shed light on the helicity distribution in the Fourier space.

In calculating the curves in Fig.~\ref{a1clas-j0}, we have used the last line of Eq.~(\ref{A1besl}) where the whole range of $P_{h\perp}$ has been integrated over. As a check, we also calculate $A_1^{J_0(|\bm{\mathcal B_T}||\bm P_{h\perp}|)}$ using the integration limit $0 \,\text{GeV}<P_{h\perp}<1.12 \,\text{GeV}$ according to the kinematical cuts at CLAS. We find that the difference of these two results is about one percent, which agrees with the conclusion in Ref.~\cite{Boer:2011xd} that the contribution of the large $P_{h\perp}$ tail is suppressed in the Bessel-weighted integrals.
This is important such that Eq.~(\ref{A1besl}) can be used in phenomenological analysis to extract $\tilde g_{1L}(x,\bm q_T^2)$ directly without worrying about
the large $P_{h\perp}$ contribution.

\begin{figure}
\begin{center}
\includegraphics[width=0.8\columnwidth]{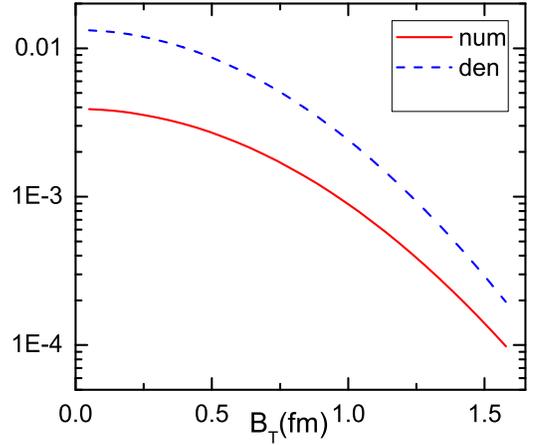}
\caption{The sizes of the numerator (solid line) and the denominator (dashed line) of Eq.~(\ref{A1besl}) as functions of $\mathcal B_T$ for $\pi^+$ production.}\label{numden1}
\end{center}
\end{figure}

In Fig.~\ref{numden1}, we plot the sizes of the numerator and the denominator of Eq.~(\ref{A1besl}) as functions of $\mathcal B_T$ for $\pi^+$ production, calculated from the light-cone diquark model.
The result shows that the measured differential cross section in the Fourier space will decrease almost exponentially with increasing $\mathcal B_T$.
Thus the region where $\mathcal B_T$ is not so large can be explored in SIDIS.
In the case the $\mathcal B_T<1.5\,\text{fm}$ region is measured,
it will provide the information of $\tilde{g}_{1L}(x,\bm b_T^2)$ in the region $b_T=z\mathcal B_T <1 \,\text{fm}$ at CLAS, where $\tilde{g}_{1L}(x,\bm b_T^2)$ should be sizable.

An interesting phenomenon is that all the Bessel-weighted asymmetries for the three pions increase with increasing $\mathcal B_T$. This can be explained by the fact that the mean square of $b_T$ for the polarized distribution in the Fourier space
\begin{align}
\langle b_T^2 \rangle_g^q =
{\int_{x_{\text min}}^{x_{\text max}} dx \int d^2\bm b_T \, b_T^2\, \tilde g_{1L}^q(x,\bm b_T^2)
\over \int_{x_{\text min}}^{x_{\text max}} dx
\int d^2\bm b_T \tilde g_{1L}^q(x,\bm b_T^2)},
\end{align}
is larger than the width $\langle b_T^2\rangle_f^q$ of the unpolarized distribution,
which is opposite to the case of TMD distributions because $b_T$ is conjugate to $p_T$. Explicitly, in Approach 2 of the light-cone diquark model we obtain $\langle b_T^2\rangle_g^{uv} =1.45\,\text{fm}^2$, $\langle b_T^2\rangle_g^{uv} =1.40 \,\text{fm}^2$,  and $\langle b_T^2\rangle_f^{q} =0.62 \,\text{fm}^2$.

\begin{figure}
\begin{center}
\includegraphics[width=\columnwidth]{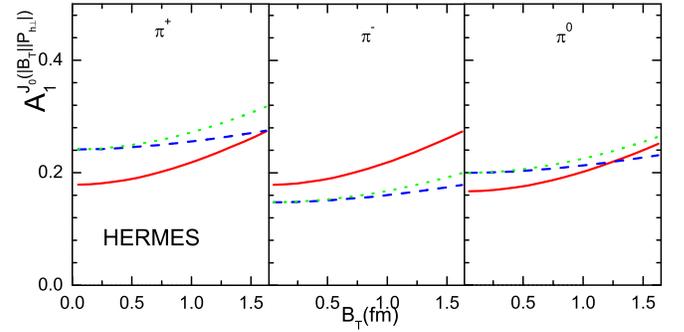}
\caption{Same as Fig.~\ref{a1clas-j0}, but for the HERMES kinematics.
}\label{a1hermes-j0}
\end{center}
\end{figure}

\begin{figure}
\begin{center}
\includegraphics[width=\columnwidth]{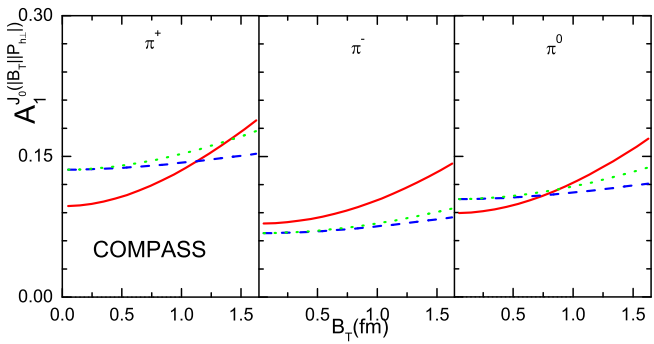}
\caption{Same as Fig.~\ref{a1clas-j0}, but for the COMPASS kinematics.
}\label{a1compass-j0}
\end{center}
\end{figure}

DSAs for meson production in SIDIS as functions of $x$ have also been measured at HERMES~\cite{Airapetian:2004zf} by
scattering a $27.6$ GeV positron beam off longitudinally polarized hydrogen and deuterium targets, as well as at COMPASS~\cite{Alekseev:2010ub} by collision of a $160$ GeV muon beam
on a longitudinally polarized NH$_3$ target.
In Figs.~\ref{a1hermes-j0} and \ref{a1compass-j0}, we present the prediction for the Bessel-weighted DSAs of $\pi^+$, $\pi^-$, and $\pi^0$ productions off a proton target in SIDIS at HERMES and COMPASS, using Approach 2 of the light-cone diquark model as well as the TMD helicity distributions from Ref.~\cite{Anselmino:2006yc} for comparison.
The kinematical cuts applied in the calculations are
\begin{align}
&0.023<x<0.6,~~
y<0.85,  \nonumber\\
&0.2<z<0.8,~~  Q^2 >1\, \textrm{GeV}^2,~~
W^2>10\,\textrm{GeV}^2\label{kine-hermes}
\end{align}
for HERMES~\cite{Airapetian:2004zf}, and
\begin{align}
&0.004<x<0.7,~~
0.1<y<0.85,  \nonumber\\
&0.02<z<1,~~  Q^2 >1\, \textrm{GeV}^2,~~W^2>25\,
\textrm{GeV}^2\label{kine-compass}
\end{align}
for COMPASS~\cite{Alekseev:2010ub}.
Again we observe that the Bessel-weighted asymmetries for the three pions  increase with increasing $\mathcal B_T$ at HERMES and COMPASS.
The difference is that the asymmetries at HERMES and COMPASS are smaller
than that at CLAS,
this is because the HERMES and COMPASS can probe the smaller $x$ region where the helicity distributions are small compared to the unpolarized TMD distributions.

\section{Conclusion}

In summary, we study the helicity distributions of valence quarks in the
transverse momentum space and the transverse coordinate space. The TMD
helicity distributions of valence $u$ and $d$ quarks are calculated
in the light-cone diquark model by adopting two different
approaches. By comparing the $p_T$ dependence of the helicity
distributions at different $x$, we find that at small $x$ and large $p_T$, the
valence $u$ quark has net negative helicity while the $d$ quark tends to have net positive helicity. We thus use our model results to analyze the $P_{h\perp}$-dependent DSAs of
$\pi^+$, $\pi^-$, and $\pi^0$ productions in SIDIS, and compare them
with the data measured by the CLAS Collaboration. We find that the
asymmetries calculated in Approach 2 of the light-cone diquark model
agree with the CLAS data. We then calculate the valence helicity
distributions in the Fourier space using Approach 2 of the
light-cone diquark model, and use them to predict the
$\mathcal{B}_T$ shape of Bessel-weighted DSAs for $\pi^+$, $\pi^-$, and $\pi^0$ productions in SIDIS at CLAS, COMPASS and HERMES for the
first time. Because the mean square transverse position of the
polarized quarks is larger than that of the unpolarized quarks, the
Bessel-weighted asymmetries for all three pions increase with
increasing $\mathcal B_T$. The contributions from large $P_{h\perp}$
region to the Bessel-weighted DSAs are calculated numerically and
found to be negligible. Our study suggests that the Bessel-weighted
asymmetry $A_1^{J_0(|\bm{\mathcal B}_T||\bm P_{h\perp}|)}$ may be
measured in the double polarized SIDIS and provides a
straightforward access to the helicity distributions in the
coordinate space.

\section*{Acknowledgements}
This work is partially supported by National Natural Science
Foundation of China (Grants No.~10905059, No.~11005018,
No.~11021092, No.~10975003, No.~11035003, and No.~11120101004),
by SRF for ROCS, SEM, and by the Fundamental Research Funds for the
Central Universities.

\end{document}